%% file: main.tex
\pdfoutput=1
\newif\ifFull
\Fulltrue

\newif\ifSim
\Simfalse

\newif\ifLineNum
\LineNumfalse

\ifFull
\documentclass [letter,11pt]{journal}
\usepackage[left=1in,top=1in,right=1in,bottom=1in]{geometry}
\usepackage{amsthm}
\else
\documentclass[10pt]{llncs}
\let\doendproof\endproof
\renewcommand\endproof{~\hfill\qed\doendproof}
\fi
\usepackage{graphicx}	
\usepackage {url,hyperref}
\usepackage {amsmath}
\usepackage {amssymb}
\usepackage {color}
\ifLineNum
\usepackage [pagewise]{lineno}
\fi
\usepackage {microtype}
\usepackage {times}

\graphicspath{{figs/}}

\newcommand {\etal}{\textit{et al.}}

\DeclareMathOperator\lca{LCA}
\DeclareMathOperator\DoSort{DO-Sort}
\newcommand{\sort}[1]{\ensuremath {\DoSort\left({#1}\right)}}

\newcommand{\cf}[1]{\texttt{#1}} 

\ifFull 
\newtheorem {theorem} {Theorem}[section]

\fi
\newtheorem{fact}[theorem]{Fact}

\ifFull
\newenvironment {repeatenv} [2]
{\vspace{5pt}\noindent{\sf #1~\ref{#2}:}\ }{}

\fi

\ifLineNum
\definecolor {infocolor} {rgb} {0.6,0.6,0.6}

\linenumbers
\fi

\definecolor {sepia} {rgb} {0.6,0.2,0.1}

\newcommand{\marrow}{\marginpar[\hfill$\longrightarrow$]{$\longleftarrow$}}
\newcommand{\beautifulremark}[3]{\textcolor{blue}{\textsc{#1 #2:}}
\textcolor{red}{\marrow\textsf{#3}}}
\renewcommand{\beautifulremark}[3]{}

\newcommand{\joe}[2][says]{\beautifulremark{Joe}{#1}{#2}}

\ifLineNum
\newcommand*\PAMEF[1]{%
\expandafter\let\csname old#1\expandafter\endcsname\csname #1\endcsname
\expandafter\let\csname oldend#1\expandafter\endcsname\csname end#1\endcsname
\renewenvironment{#1}{\linenomath\csname old#1\endcsname}{\csname oldend#1\endcsname\endlinenomath}}%
\newcommand*\PBAMEF[1]{\PAMEF{#1}\PAMEF{#1*}}
\AtBeginDocument{\PBAMEF{equation}\PBAMEF{align}\PBAMEF{flalign}\PBAMEF{alignat}\PBAMEF{gather}\PBAMEF{multline}}
\fi

\title { Data-Oblivious Graph Algorithms \\ in Outsourced External Memory
}
\ifFull
\shorttitle {Data-Oblivious Graph Algorithms}
\else
\titlerunning{Data-Oblivious Graph Algorithms}
\fi

\author
{  
Michael T. Goodrich
\ifFull
and
\else
\and
\fi
Joseph A. Simons
\ifFull\\
Dept. of Computer Science, Univ. of California, Irvine, USA
\fi
}
\ifFull
\shortauthor{Goodrich and Simons}
\else
\authorrunning{Goodrich and Simons}
\fi

\ifFull
\else
\institute{ 
Dept. of Comp. Sci., Univ. of California, Irvine, USA;
\texttt{\{goodrich,jsimons\}@uci.edu}}
\fi
\begin {document}
\pagestyle{plain}
\maketitle

\begin {abstract}
Motivated by privacy preservation for outsourced data,
\emph{data-oblivious external memory} is a computational framework
where a client performs computations on data
stored at a semi-trusted server in a way that does not reveal her data
to the server.
\ifFull
This approach facilitates collaboration and reliability 
over traditional frameworks, and it provides privacy protection, even though
the server has full access to the data and he can
monitor how it is accessed by the client. 
The challenge is that even if data is encrypted, 
the server can learn information based on the client data access pattern;
hence, access patterns must also be obfuscated.
We investigate privacy-preserving algorithms
for outsourced external memory
that are based on the use of data-oblivious algorithms, that is,
algorithms where each possible sequence of data accesses 
is independent of the data values.
\fi
We give new efficient data-oblivious algorithms in the outsourced 
external memory model for a number of fundamental graph problems. 
\ifFull
Our results include
new data-oblivious external-memory methods for
constructing minimum spanning trees,
performing various traversals on rooted trees,
answering least common ancestor queries on trees,
computing biconnected components,
and forming open ear decompositions.
None of our algorithms make use of constant-time random oracles.
\fi
\end {abstract}

\input{intro.tex}

\section{Tree-Traversal Computations}
\label{sec:tree-comp}
\label{sec:euler-tour}
Many traditional graph algorithms are based on a traversal of a spanning tree of
the graph, for example, using depth first search. 
However, the data access pattern of depth first search fundamentally depends on
the structure of the graph, and it is not clear how to perform DFS efficiently
in the DO-OEM model. 
Instead, we use
\emph{Euler Tours}~\cite{tv-epba-85}, adapted for data-oblivious tree-traversal
computation~\cite{got-gdc-12}.
Given an undirected rooted tree $T$, we imagine that each edge $\{p(v), v\}$ is
composed of two directed edges $(p(v), v)$ and $(v, p(v))$, called an
\emph{advance} edge and \emph{retreat} edge respectively. An \emph{Euler tour}
of $T$ visits these directed edges in the same order as they would be visited in
a depth first search of $T$.
However, an Euler tour implemented with compressed-scanning does not reveal
information to the adversary because each data item is accessed once and in
random order.

\ifFull
 We give more details concerning the implementation
of an Euler tour in Appendix~\ref{app:euler-tour}.
Let \cf{E-order}$(u,v)$ denote the order of the
edge $(u,v)$ in an \emph{Euler tour} of $T$. 
Note that 
$\cf{preorder}(v) = \cf{E-order}(p(v),v)$ and 
$\cf{postorder}(v) = \cf{E-order}(v,p(v))$. 
\fi

Some tree statistics are straightforward to compute
using Euler Tours. For example, Goodrich~\etal~\cite{got-gdc-12} show how to
compute the size of the subtree for each node $v \in T$ using an Euler Tour
and compressed-scanning pass over the edges of $T$. 
The calculation is straightforward once we observe that
$\cf{size}(v) =  (\cf{E-order}(v,p(v)) - \cf{E-order}(p(v), v))/2 + 1$,
since for each proper descendant of $v$, we will traverse one advance edge and
one retreat edge. Thus, the number of edges traversed between $(p(v), v)$ and
$(v, p(v))$ is twice the number of proper descendants of $v$, and we add one to
also include $v$ in $\cf{size}(v)$. 
Moreover,  
Euler tour construction 
can be done data-obliviously
in external memory
in $O(\sort{|T|})$ I/Os by a data-oblivious compressed-scanning
implementation of the algorithm by Chiang {\it et al.}~\cite{Chiang:1995}.

However,
Euler-Tours are insufficient to compute most functions where the value at a
vertex is dependent on its parent or children. Therefore, in the following, we
describe more sophisticated techniques for tree \emph{traversal computations}, suitable
for computing functions in which the value at a vertex depends on its parent or
children.

\ifFull
\subsection{Bottom-Up Computation.}
\else
\paragraph{Bottom-Up Computation.}
\fi
Let $T$ be a tree rooted at $r$.
First, we show how to compute recursive functions on the vertices bottom up
using a novel data-oblivious algorithm inspired by the classic parallel tree contraction
of Miller and Reif~\cite{mr-ptc-85}. 
Like Miller and Reif's algorithm, 
our algorithm compresses a tree down to a single node in $O(\log V)$ rounds.
However, unlike the original algorithm, we are able to compress long paths
of degree two nodes into a single edge in a single iteration and 
guarantee the size of the graph decreases by half in each round.

Each round of the tree contraction algorithm is divided into two operations: \cf{rake}, which
removes all the leaves from $T$, and
\cf{compress}, which compresses long paths by contracting edges for which the
parent node only has a single child. 

First, we label each vertex with its degree by scanning 
the edge list in adjacency list order. Then, for
each vertex $v \neq r$, if $v$ has degree 1, it is a leaf, and it is marked for
removal by the rake operation. Otherwise, if it has degree 2, then it is marked
for contraction by the compress operation. 
The marks are stored with the endpoints of each edge.

Now, we perform the \cf{rake} operation via an Euler tour of $T$. For each
unmarked edge we read, we write back its value unchanged.  If an advance edge is
marked as a leaf, then we mark the edge for removal.
The next edge we read is the corresponding retreat edge from the leaf. We
evaluate the leaf and  output the computed value together
with the label of the parent vertex instead of the original retreat edge. 
Next we distribute this information to the other incident edges;
we sort the edge list so that for each vertex we first see all
the evaluated leaves and then see the remaining outgoing edges.  In a compressed
scanning pass we are able to store the function evaluation from each leaf in
its parent. 
Thus we complete the \cf{rake}
operation.

\begin{figure}[hbt]

\begin{minipage}[t]{.45\linewidth}
\centering
\includegraphics[width=\textwidth]{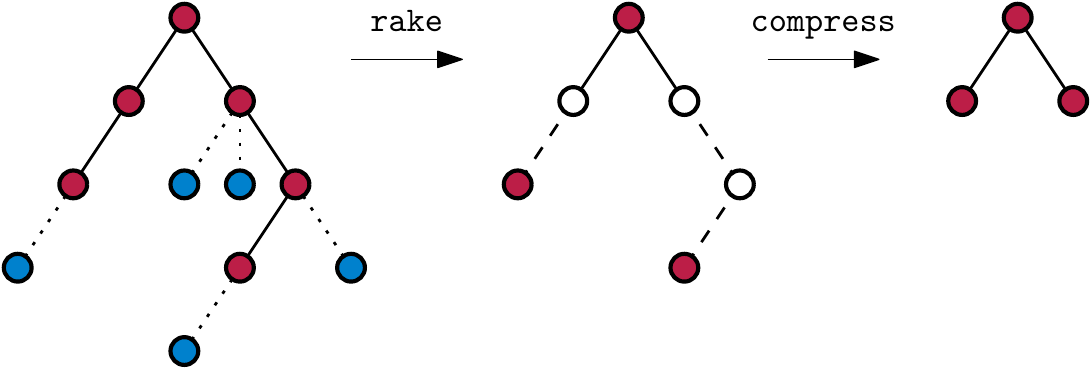}
\caption{A single round of the \cf{rake} and \cf{compress} operations on a small
example graph.}
\label{fig:rake-compress}
\end{minipage}
\hfill
\begin{minipage}[t]{0.45\linewidth}
\centering
\includegraphics[width=\textwidth]{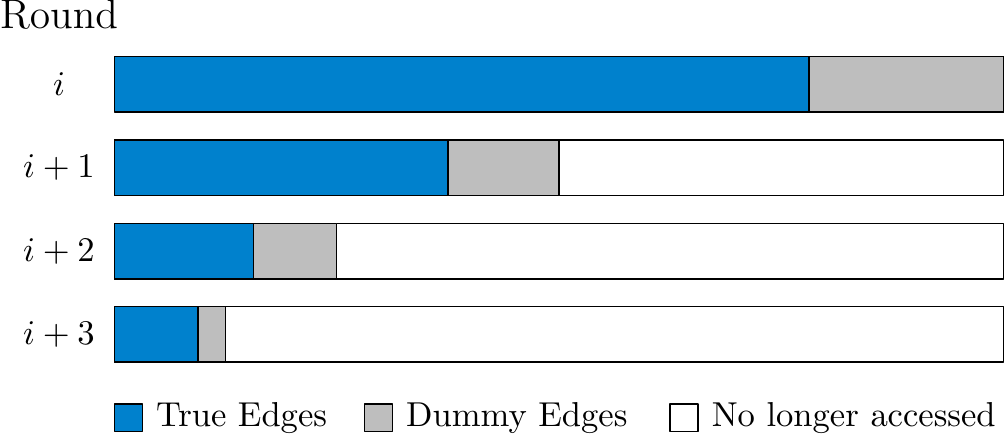}
\caption{
    The portion of memory accessed by the algorithm decreases by a constant
    fraction in each round.
    }
\label{fig:geom-sum}
\end{minipage}

\end{figure}


Next, we perform the \cf{compress} operation via another Euler tour of $T$. 
We remove each marked advance edge by writing dummy values in its place. For
each retreat edge to a degree two node, we contract the edge by composing the
functions at the parent and child and storing this value in private memory. For
all but the last edge in a path of degree two nodes, we mark the edge for
removal. For the last edge in the path, we output the label of the
parent vertex together with the composition of all functions along the path.
Although the path may not have constant size, for the functions considered in
this paper (such as min), the composition across values of nodes along the path
can be expressed in $O(1)$ space by partially evaluating the function as we go. 
We  pass this information to other edges
incident to the last vertex in the path, for all the compressed paths, 
using a single compressed scanning round.
Thus, we complete the \cf{compress} operation.
(See Figure~\ref{fig:rake-compress}.)

Finally, we perform one last compressed scanning pass to set aside all edges marked
for removal. These edges are placed at the end of the list 
by the sort and are not required for subsequent processing.
Thus we complete one round of
the algorithm.
However,
we may continue to access some dummy edges in subsequent rounds so that in round $i+1$ we always
access a constant fraction of the memory accessed in round $i$, thus maintaining
the data-oblivious property of our algorithm (see 
Figure~\ref{fig:geom-sum}). 

We now analyze the total time it takes to contract a tree down to the root node.
Each round of \cf{rake} and \cf{compress} on a tree $T_i$ of size $V_i$ takes 
$O(\sort{V_i})$ time to
perform $O(1)$ Euler tours and compressed scanning rounds. 
We begin with an initial tree $T_0 = T$ of size $V_0 = V$. Without loss of
generality, we can partition the nodes of any tree $T_i$ into three sets: 
$B_i$, the branch nodes with at least $2$
children; $P_i$, the path nodes with $1$ child; and $L_i$, the leaf nodes.
Clearly $|B_i| + |P_i| + |L_i| = V_i$, and $|L_i| \geq 2|B_i|$. The rake
operation removes all of $L_i$, and all but at most one node in each path in
$P_i$. Thus, $V_{i+1} \leq |B_i| + \frac{1}{2}|P_i| 
< \frac{1}{2} (3|B_i| + |P_i|) \leq \frac{1}{2}V_{i}$. 
Hence, $\sum_i V_i$ is a geometric sum, and the total running time of all
rounds is $O(\sort{V})$.

Throughout the algorithm the children of each branch and leaf node are finalized
before we process the node. 
However, some path nodes may have been compressed and set
aside before all of their descendants were finalized.
Thus, in a final post-processing step, we perform one more round of compressed
scanning and Euler tour over the full edge list to finalize the value of the 
internal path nodes.
 
\paragraph{Computation of \cf{low}.}
\ifFull
\label{sec:low}
\else
\label{app:low}
\fi
Suppose we are given a spanning tree $T$ of a graph $G$.
We illustrate rake and compress by computing the following simple recursive
function.
\begin{align*}
 	\cf{low}(v) = \min(&\{\cf{preorder}(v))\} \\
    &\cup 
	\{\cf{low}(w) \mid \mbox{ $w$ is a child of $v$ in $T$} \} \\
    &\cup
	\{ \cf{preorder}(w) \mid (v,w) \in G - T \})
\end{align*}
That is, for each vertex $v \in T$, 
$\cf{low}(v)$ is the lowest preorder number of a vertex that is a descendant of
$v$ in $T$, or adjacent to a descendant via a non-tree edge. 
This function is a key part of the biconnected
components algorithm, and key functions in our other algorithms are 
computed similarly.

First, we compute the preorder numbers of each vertex by an Euler tour of $T$.
Next, we preprocess the edge list. In $O(1)$ compressed-scanning rounds, we
compute for each vertex the minimum preorder number between that vertex and all 
its neighbors in $G - T$. 
We store this data in the endpoints of each edge $(u, v)$ in the edge list as the
initial values for $\cf{low}(u)$ and $\cf{low}(v)$. This preprocessing requires
$O(\sort{E})$ time.

Now, we use rake and compress to compute the recursive portion of $\cf{low}$.
Each iteration of rake and compress proceeds as follows.
The low value of each leaf is already finalized. 
We store
$ \cf{low}(v) = \min\left( \{\cf{low}(v)\} \cup \{\cf{low}(w) \mid
w \text{ is a child of }v  \} \right)$ as the function for each internal node
$v$. When we rake
a leaf $\ell$, we update its parent $p$, $\cf{low}(p) =
\min(\cf{low}(p), \cf{low}(\ell)).$ 
During the compress step, when we contract an edge $(p(v), v)$ we update the
function stored at in $p(v)$ as follows:
$\cf{low}(p(v)) = \min(\cf{low}(p(v)), \cf{low}(v))$, and
$\cf{children}(p(v)) = \cf{children}(v)$. Note that vertex may have many
children, but always a single parent. Thus, the easiest way to change the
assignment of children is to set $\cf{label}(p(v)) = \cf{label}(v)$ and then
relabel the edge $(p(p(v)), p(v))$ = $(p(p(v)), v)$. We may need to perform this
relabeling and calculation of $\cf{low}$ over a long path, 
but 
we always process nodes bottom up, 
and we maintain the values of the previous edge
processed in private memory. We output dummy values for all but the
final edge in the path, which stores the minimum \cf{low} of the whole path,
together with the labels of the first and last vertex on the path.
Finally, we synchronize each edge with the new values of its
endpoints via compressed-scanning, which completes the iteration of rake and
compress. After at most $O(\log V)$ iterations and $O(\sort{V})$ I/Os, we
complete the rake and compress algorithm. Thus, the preprocessing time dominates, and
the total time required to compute $\cf{low}$ for all vertices is $O(\sort{E})$.

\ifFull
\subsection{Top-down computation}
\else
\paragraph{Top-Down Computation.}
\fi
\label{sec:top-down}
We now show how to run our compression algorithm ``in reverse" in
order to efficiently compute top-down functions 
where each vertex depends on
the value of its parent. First, we simulate the compression algorithm described
above, and label each edge $e$ with \cf{contract}$(e)$, 
the order in which it would have been removed
from the graph. Thus, all edges incident to leaves in the initial graph are given
a label smaller than any interior nodes, and all edges incident to the root are
given larger labels than edges incident to nodes of depth~$> 1$. 

Next, we sort the edges in reverse order according to their \cf{contract} labels. 
We process the edges in this order in $\lceil \log V\rceil$ stages.  
We mark the root as \emph{finished}.
Then, in each stage $i$, we perform the following on the 
first $2^i$ edges in the sorted order:
For each advance edge $e = (p(v), v)$, if $p(v)$ is marked as \emph{finished}, we
evaluate the function at $v$, augment $e$ with its value, and mark $v$ as
\emph{reached}. Between stages $i$ and $i+1$, we process the first $2^{i+1}$
edges, and distribute the new values at reached vertices from the previous stage
to any incident edges belonging to the next stage. 
Finally, we mark each reached vertex as finished. 
Thus, the function at a parent is always evaluated before the function at its
children, and each child edge has been augmented with the value from the parent
before the edge is processed. 

Each stage requires $O(1)$ rounds of compressed scanning.
Since the number of edges processed in each stage is $2^i$, 
the running time of the final stage dominates all other
stages, and thus the total time is $O(\sort{V})$.
Note that since our algorithm essentially reduces to data-oblivious sorting and compressed
scanning, the sequence of data accesses made by the algorithm are independent of
the input values. Thus, no probabilistic polynomial adversary has more than
negligible advantage in the input-indistinguishability game. 
We summarize our results in the following theorem:

\begin{theorem}
Given a tree $T$, we can perform any top-down or bottom-up 
tree-traversal computation over $T$ in $O(\sort{V})$ time in the DO-OEM model.
\end{theorem}

\paragraph{LCA computation.} \label{sec:LCA}
Suppose we are given a connected graph $G=(V,E)$ and a spanning tree $T=(V, E_T)$ of $G$
rooted at $t$.
We can preprocess $G$ and augment each edge 
$(u,v) \in E$ with additional information such that we can find the
least common ancestor 
$\lca(u,v)$ with respect to $T$ in constant time. 
Given two integers $x,y$, let $\cf{rzb}(x)$ denote the number of rightmost zero bits in the
binary representation of $x$, and  let $x \& y$ denote the
bitwise logical AND of $x$ and $y$. 
The following preprocessing algorithm is adapted from the 
parallel algorithm of Schieber and Vishkin~\cite{sv-flca-88}. 

For each node $v \in T$, we compute \cf{preorder}$(v)$, and \cf{size}$(v)$. 
We also set $\cf{inlabel}(v)$ to $\max_{w \in T_v} \cf{rzb}(\cf{preorder}(w))$,
that is, the maximal number of rightmost zero bits of any of the preorder numbers of the
vertices in the subtree rooted at $v$.
As in our computation of $\cf{low}$ in 
\ifFull
Section~\ref{sec:low}, 
\else
Section~\ref{app:low},
\fi{}
we can compute these functions via rake and compress together with a 
$O(1)$ Euler tour and compressed-scanning
steps so that each edge $(u,v) \in G$ stores the
augmented information associated with its endpoints.
We also initialize
$\cf{ascendent}(t) = 2^{\lfloor \log V \rfloor}$.

We  compute $\cf{ascendant}(v)$ for each vertex
$v \in V$ as follows:
If $\cf{inlabel}(v)$ is equal to $\cf{inlabel}(p(v))$, then
set $\cf{ascendent}(v)$  to $\cf{ascendent}(p(v))$. 
Otherwise, set
 $\cf{ascendent}(v)$ to $\cf{ascendent}(p(v)) + 2^i$, where 
 $i = \log(\cf{inlabel}(v) - [\cf{inlabel}(v) \& (\cf{inlabel}(v) - 1)])$ 
is the index of the rightmost non-zero bit in $\cf{inlabel}(v)$.
Thus, $\cf{ascendent}$ is a top-down function, and we can evaluate it at all
nodes in the tree in $O(\sort{V})$ time using the method of
Section~\ref{sec:top-down}.

Finally, we perform an Euler-Tour traversal to compute a table $\cf{head}$.
We set 
$\cf{head}(\cf{inlabel}(v))$ to be the vertex of minimum depth $d(u)$ among all the
vertices $u$ such that $\cf{inlabel}(u) = \cf{inlabel}(v)$ 
on the path from the $t$ to $v$ in $T$.
Note that since there are at most $\log V$ distinct $\cf{inlabel}$ numbers, the size of
\cf{head} is at most $O(\log V)$ and can fit in private memory. 
By a constant number of compressed-scanning steps, we store $\cf{head}(\cf{inlabel}(u))$
and $\cf{head}(\cf{inlabel}(v))$ with each edge $(u,v) \in G$.

Given the additional information now stored in each edge, we can compute the
$\lca(u,v)$ for any edge $(u,v) \in G$ in constant time by a few simple algebraic
computations as shown by Schieber and Vishkin~\cite{sv-flca-88}. 

We summarize this result in the following theorem:
\begin{theorem}
    \label{cor:LCA}
Given a tree $T$ and a set of pairs of vertices $S \subset V \times V$, 
we can compute $\lca(u,v)$ for all $u,v \in S$ in $O(\sort{|S| + V})$ 
time in the DO-OEM model.
\end{theorem}

%
\ifFull
\ifSim
\input{sim}
\fi
\fi
\ifFull
\section{Minimum Spanning Tree}
\else
\section{Algorithms}
\fi

In this section we present a novel algorithm to compute the 
minimum spanning tree of a general graph in the DO-OEM model. Our algorithm has
additional input parameters of the density and class of the graph, and its
running time depends on these parameters. 
Thus, our algorithm necessarily reveals asymptotically the vertex and edge counts, and
whether the input graph is minor-closed.
However, revealing this information does not convey any advantage to the
adversary in the input-indistinguishability game. 
In fact, these input parameters can be freely chosen by the adversary. 
Of course, if we don't want to allow these input parameters and the
corresponding gains in efficiency, 
we can avoid revealing this information by working with an adjacency
matrix instead of an edge list, or we can also achieve a
tradeoff between privacy and efficiency by padding the input with dummy edges.

%
In the case of somewhat dense graphs, or graphs from a minor closed family, our
runtime is $O(\sort{E})$. We conjecture that this is optimal since we require this time
to perform even a single round of compressed scanning in our model. 
For graphs of other classes and densities, our algorithm still beats the
previously best known method for computing \emph{any} spanning tree in the
DO-OEM model by logarithmic factors.

\begin{figure}[hbt]
\vspace*{-16pt}
\centering
\includegraphics[width=.85\textwidth]{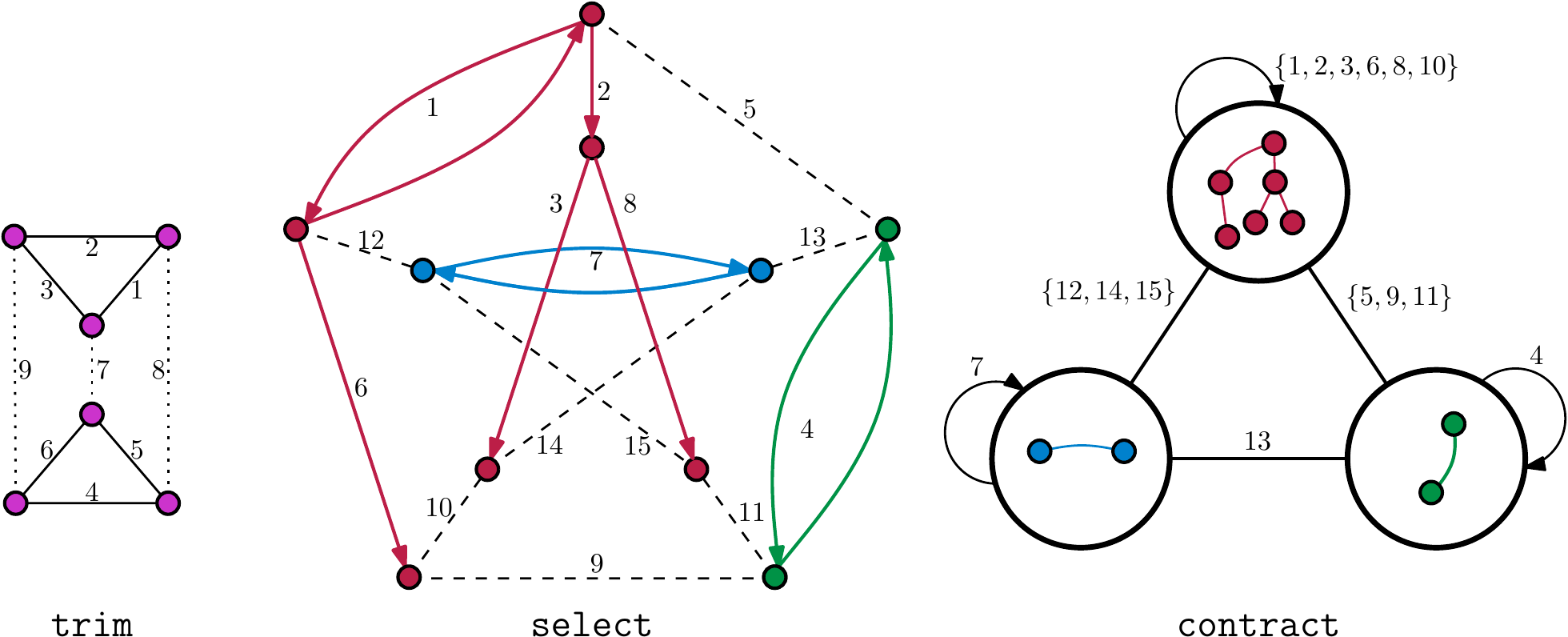}
\vspace*{-4pt}
\caption{Left: the result of \cf{trim} on a small example graph with $\alpha = 2$.
Removed edges are denoted by dotted lines.  
Center and Right: the result of \cf{select} and \cf{contract} on a small example graph. 
In \cf{select}, unselected edges
are dashed, and selected edges are solid and oriented towards the vertex which
selected them. The selected edges form a spanning forest of $G$; 
each connected component is a tree in which one edge has been selected twice. 
In \cf{contract}, each connected component is contracted to a single
pseudo-node.}
\label{fig:trim}
\label{fig:select}
\end{figure}

\ifFull
\else
We also leverage our minimum spanning tree and tree-traversal computation
algorithms to design new data-oblivious algorithms to compute biconnected
components of a graph, and an open ear decomposition and st-numbering of a
biconnected graph. 
\subsubsection{Minimum Spanning Tree}
\fi
\label{sec:mst}
%
Our MST algorithm requires the following sub-routines: 
\cf{trim}, \cf{select}, \cf{contract} and \cf{cleanup}.
\begin{description}
\item \cf{trim}$(G, \alpha)$: \\
We scan the edge list of $G$ and trim the outgoing edges from
each node, depending on the value of an input parameter $\alpha$. 
If the degree of a node is 
at most $\alpha$, then we leave its outgoing edges unchanged.
If the degree of a node is less than $\alpha$, then we implicitly 
pad its outgoing edge list up to
size $\alpha$ with
additional dummy edges of weight $\infty$. 
However, if the degree of a node is greater than $\alpha$, then the node keeps its $\alpha$
smallest outgoing edges and discards the rest. For a given edge
each endpoint independently chooses to keep or discard it as an outgoing edge. 
Note that we can trim
the edge list in $O(1)$ rounds of compressed scanning. Afterwards, the
graph may no longer be connected (see Figure~\ref{fig:trim}).

\item \cf{select}$(G)$: \\ 
Each node selects the minimum outgoing edge from its adjacency list. If two edges have the
same weight, we break ties lexicographically. 
We mark the edges as selected as follows:
First, sort the edges lexicographically by source vertex, weight. 
Then, in a single compressed scanning round we mark the minimum weight edge from
each source vertex as selected. The set of selected edges partition $G$ into connected
components, and in each connected component, one edge has been selected twice
(see Figure~\ref{fig:select}). For each double-selected edge, we arbitrarily choose to
keep one copy and mark the other as dummy. We can gather all the dummy edges
to the end of the list
in $O(1)$
compressed scanning rounds. The output is a spanning forest of $G$ such that
in each tree all the edges are oriented from the root to the leaves.


\item \cf{contract}$(G, E_s)$: \\
The input is a graph $G$ and a set of selected edges $E_s$ which induce a
spanning forest $F \subseteq G$.
For each connected component in $F$, we
merge the nodes in the component into a single pseudo-node 
by contracting all the selected edges
in that component (Figure~\ref{fig:select}).
Using our top-down tree-traversal algorithm, we re-label each node with the
label of the root of its component. 
Then, in two compressed scanning rounds we relabel the endpoints of the all edges 
in $G$ to reflect the relabeled nodes, possibly creating loops and parallel edges. 
In the first round we relabel the source for each outgoing edge
of each node. In the second round we sort the edges to group them by 
incoming edges with each node, and relabel the target for each edge. 
In a final oblivious sort, we restore the edge list of $G$ to adjacency list 
(lexicographical) order.


\item \cf{cleanup}$(G)$: \\
We detect and remove duplicate, parallel and loop edges in a single compressed-scanning round.
When we encounter parallel edges, we remove all but the minimum weight edge between two
nodes. As we scan the edge list, we remove an edge by writing a dummy value in
its place, and then we perform a final oblivious sort to place all the dummy values
at the end of the list.

\end{description}


%

 \paragraph{Minimum Spanning Forest (MSF).} 
Let $G_0 = \cf{trim}(G, \alpha)$ for an appropriate choice of parameter $\alpha$ to
be discussed later.
Then,
 the core of our algorithm is as follows:
for $i \in [1, \alpha]$, let 
\[
G_i = \cf{trim}(\cf{cleanup}(\cf{contract}(G_{i-1},\cf{select}(G_{i-1}))), \alpha).
\]
That is, we perform $\alpha$ iterations in which we select the minimum edge out of each
node, contract the connected components, ``remove'' unwanted edges from the
resulting graph by labeling them as dummies,
and pass the cleaned and trimmed graph to the next iteration. Each subsequent
iteration accesses a constant fraction of the memory accessed in the previous
iteration, possibly including some dummy edges (see Figure~\ref{fig:geom-sum}).
In a final pass, we \cf{contract} and \cf{cleanup} 
all the edges of $G$ with respect to the connected components
represented by the nodes of $G_\alpha$.

Throughout our algorithm, the set of selected edges induce a spanning forest of
$G$. Each pseudo-node represents an entire tree in this forest. 
We define a weight function $w(v)$ for each pseudo-node, which corresponds to
the number of true nodes contained in the tree represented by the pseudo-node.
Initially each node has weight 1.

At each \cf{trim} step, we remove a subset of edges. The remaining edges induce
a set of \emph{potential} connected components $C_1, \ldots, C_t$.
For each $C_i$, let $w(C_i) = \sum_{v \in C_i} w(v)$.
We maintain the following invariant throughout all iterations of our algorithm:
$w(C_i) \geq \alpha$ for all $C_i$.

The invariant remains true after the initial \cf{trim} step;
we know that each $C_i$ contains at
least $\alpha + 1$ nodes, since $G$ was connected, and \cf{trim} only removes
edges from nodes of degree more than $\alpha$. Subsequent \cf{trim} steps also
maintain the invariant since nodes of degree~$\leq \alpha$ are not effected and
nodes of degree $> \alpha$ will still be connected to at least $\alpha$ other
nodes after the trim.

Next, in the $\cf{select}$ step, each node selects one outgoing edge for
contraction. For each edge $(u,v)$ that we contract in the \cf{contract} step, 
we create a new pseudo-node $x$ of weight $w(x) = w(u) + w(v)$. 
Thus, the total weight of each $C_i$ does
not change over the contract step. However, we select at least $|C_i|/2$ edges
in each component. 
Thus, the number of nodes within each component $C_i$ is reduced by half.

The \cf{cleanup} step only removes redundant edges, and does not effect the number of
nodes or the weight in any components.

Hence, after $O(\log \alpha)$ iterations, the size of each $C_i$ is reduced by a
factor of $\alpha$, and the total weight of each $C_i$ remains at least $\alpha$. 
Therefore, the resulting graph has at most $O(V / \alpha)$ pseudo-nodes.

We now consider the running time of the MSF algorithm.
The first run of \cf{trim}$(G)$ and the final pass both take $O(\sort{E})$ time.
Each sub-routine used in each iteration takes 
$O(\sort{|G_i|})$ time.
Moreover, for each $i$, 
$|G_{i+1}| \leq |G_i| / 2$.
Therefore, the time used in all iterations is a geometric sum, and the time required
by all iterations is $O(\sort{|G_0|}) = O(\sort{\alpha \cdot V})$.
Hence, the total running time of one repetition of MSF is 
$O(\sort{E} + \sort{\alpha \cdot V})$.
Moreover, we can repeat our algorithm $k$ times to achieve the following result.
Let $G^j(\alpha)$ denote the graph for the 
$j$th repetition of the MSF algorithm with parameter $\alpha$, that is, the
graph consisting of the pseudo-nodes (subsets of vertices of $G$) 
from the output of the previous
iteration, and the $\alpha$ smallest edges out of each pseudo-node.
The running time of $k$ repetitions is
\ifFull
\begin{align*}
    & \sum_{j=0}^{k-1} O(\sort{E} + \sort{\alpha |G^j(\alpha)|}) \\
    &= O(k \cdot\sort{E}) + \sum_{j = 0}^{k-1} O(\sort{\alpha \frac{V}{\alpha^j}}) \\ 
    &= O(k\cdot \sort{E} + \sort{\alpha V})
\end{align*}
\else
\begin{align*}
     \sum_{j=0}^{k-1} O(\sort{E} + \sort{\alpha |G^j(\alpha)|}) 
    &= O(k\cdot \sort{E} + \sort{\alpha V})
\end{align*}
since $\sum |G^j(\alpha)|$ is a geometric sum.
\fi
%
%
%
%
We require that the final output
$|G^k(\alpha)| = 1$ is a single pseudo-node representing a spanning tree of $G$, 
which implies that $k = \frac{\log V}{\log \alpha}$. 
Thus, we choose parameter 
\ifFull
\[
\alpha = \frac{\beta}{\log \beta}\mbox{, where }  \beta = \frac{E \log V} { V} 
\]
\else
$\alpha = \frac{\beta}{\log \beta}$, where  
$\beta = \frac{E \log V} { V}$ 
\fi
to minimize the total running time, depending
on the density of the graph. Hence, the running time is
%
\[
    O\left( 
        \sort{E} \frac{\log V}{\log \frac{\beta}{ \log\beta}} + 
        \sort{E\frac{\log V}{ \log \beta}}  \right)
\]
%
If the graph $G$ is somewhat dense with $E = \Omega(V^{1+\epsilon})$ for any
arbitrarily small constant $\epsilon > 0$, then this implies a running time of 
$O(\sort{E})$. 
If $G$ has density $E = \theta(V2^{\log^\delta V})$, for any constant 
$0 <\delta < 1$ then this implies a running time of 
$O(\sort{E}\cdot \log^{1-\delta}V)$.
If $G$ is sparse with $E = O(V \log^\gamma V)$, for any constant $\gamma \geq 0$,
then we achieve a running time of $O(\sort{E}\cdot \log V / \log \log V)$. 
However, if $G$ is from a minor closed family, e.g. if it is a planar
graph,
we can be even more efficient and simplify our algorithm.

Mare{\v{s}}~\cite{m-tltamst-04} gave a related algorithm in the standard RAM model (not
data-oblivious). He showed that for any non-trivial minor closed 
families of graphs, a Bor\r{u}vka-style round of edge contractions will always decrease
the size of the graph by a constant factor. 
Therefore, we will have a similar geometric sum
in the running time of our algorithm 
for any input graph drawn from a non-trivial minor closed family of
graphs, including any graph with bounded genus.

Thus, in this case we do not need the \cf{trim} 
sub-routine at all, since the size of the graph will be decreasing geometrically
by the minor closed property. 
Then we run our algorithm with parameters $\alpha = E$ and $k = 1$. 
Therefore, the total time required for all iterations is $O(\sort{E})$.
Furthermore, since we always selected the minimum edge out of each component,
by the cut property of minimum spanning trees we are
guaranteed that the algorithm produces a minimum spanning tree of $G$. 
Moreover,  The algorithm makes choices depending on the input
parameters but the fundamental components of our algorithm are data-oblivious sorting
and compressed scanning and the memory access pattern never depends on the data values. 
Thus, no probabilistic polynomial
time adversary has more than negligible advantage in the input-indistinguishability
game.

We summarize our result in the following theorem:
\begin{theorem}
In the DO-OEM model, we can construct a minimum spanning tree of a graph $G$
 in time depending on the input parameters $V$, $E$, density, and class of $G$. 
 If $G$ belongs to
a minor-closed family of graphs, such as any graph with bounded genus, then the
running time is $O(\sort{E})$. We achieve the following run-times:
\begin{center}
\begin{tabular}{|c|c|c|c|}
\hline
Density & Class & Running Time & Constants \\  \hline
$E = O(V \log^\gamma V)$ & Any & $O(\sort{E} \log  V / \log \log V)$ & $\gamma \geq 0$ \\ \hline
$E = \Theta(V2^{\log^\delta V})$ & Any & $O(\sort{E} \log^{1-\delta}V)$ & $0 < \delta < 1 $  \\\hline
$E = \Omega(V^{1+\epsilon})$ & Any &  $O(\sort{E})$ & $0 < \epsilon \leq 1 $ \\ \hline
Any & Minor Closed & $O(\sort{E})$ & --- \\\hline
\end{tabular}
\end{center}
\end{theorem}

Given the minimum spanning tree algorithm and tree-traversal computation
technique outlined above, we can also achieve the following results for a
biconnected graph. 
\ifFull
We can verify that a graph is biconnected or compute the
biconnected components of a graph using the techniques
described in Section~\ref{sec:biconnected}.
\fi

\ifFull
\section{Biconnected Components}
\label{sec:biconnected}
Tarjan and Vishkin~\cite{tv-epba-85} gave an efficient sequential algorithm for
computing the biconnected components of a graph. We rely on the correctness
of their algorithm, but the details of each step are necessarily
different in our data-oblivious biconnected components algorithm.

Suppose we are given a spanning tree $T$ of a graph $G$.
Using our traversal computation method, we compute $\cf{low}(v)$ for each vertex
$v \in T$ as outlined in Section~\ref{sec:low}. We also compute a similar
function $\cf{high}(v)$, using identical techniques. The function $\cf{high}$ is
the same as $\cf{low}$ except we recursively compute the $\max$ instead of 
$\min$ preorder number. We also compute for each vertex its preorder number, 
the size of its subtree and its grandparent in $T$ (if it exists).

After this preprocessing, we construct the following edge list of an auxiliary 
graph $G''$. Each vertex of $G''$ corresponds to an edge of $T$. We add edges to
$G''$ as follows:
\begin{itemize}
\item
For each edge $\{p(v), v\} \in T$, such that $\cf{preorder}(v) \neq 1$, and
\\ either
$\cf{low}(v) \leq \cf{preorder}(p(v))$  or 
$\cf{high}(v) \geq p(v) + \cf{size}(p(v))$, 
\\we add the edge 
$\{\{p(p(v)), p(v)\}, \{ p(v), v\} \}$ to $G''$.
\item
For each edge $\{v,w\} \in G - T$ such that \\
$\cf{preorder}(v) + \cf{size}(v) \leq \cf{preoder}(w)$, \\ we add the edge
$\{\{p(v), v\}, \{p(w), w \}  \}$ to $G''$.
\end{itemize}
Now, we find the connected components of $G''$ 
and label each edge of $T$ with the label of its
component in $G''$ using the top-down tree-traversal computation technique of
Section~\ref{sec:top-down}.
Next, give each edge $\{v, w\}$ of $G-T$ with $\cf{preorder}(v) <
\cf{preorder}(w)$ the label of edge $\{p(w), w\}$ in $T$ in $O(1)$ compressed
scanning passes. 
Tarjan and Vishkin~\cite{tv-epba-85} showed the resulting component labels
correspond exactly to the biconnected components of $G$. Thus, we can compute
the biconnected components of $G$ in $O(\sort{E})$ time. We summarize this
result in the following theorem:
\begin{theorem}
Given a graph $G$ and a spanning tree of $G$, we can compute the biconnected
components of $G$ in $O(\sort{E})$ time in the DO-OEM model.
\end{theorem}
\fi{}
\ifFull{}
\else
\begin{theorem}[Biconnected Components]
Given a graph $G$ and a spanning tree, we can compute the biconnected
components of $G$ in $O(\sort{E})$ time in the DO-OEM model.
\end{theorem}
\fi

\ifFull
\section{Open Ear Decomposition}
\else
\subsubsection{Open Ear Decomposition}
\fi
\label{sec:ear-decomp}
Let $T = (V,E_T)$ be a spanning tree of $G$ rooted at $t$ 
such that there is only a single edge $(s,t) \in E_T$ incident to $t$.
The edges of $E - E_T$ are denoted \emph{non-tree edges}, and the edges $E_T - (s,t)$ are
denoted \emph{tree edges}. The edge $(s,t)$ is treated separately.
Let $G$ be a biconnected graph. 

We structure our algorithm around the same steps used by Maon~\etal~\cite{msv-pedsstn-86}
in their parallel ear-decomposition algorithm. 
However, the details of each step are
necessarily different in important ways as we show how to efficiently implement each step 
in the DO-OEM model.

\begin{enumerate}
\item Find a spanning $T$ of $G$ rooted at $t$ such that $(s,t)$ 
is the only edge incident to $t$. 
We remove the vertex $t$ from $G$ and run the above MST algorithm on $G[V - \{t\}]$ to get
a set of edges $E'$  which form a spanning tree  of $V - \{t\}$. 
Let $E_T = E' \cup (s,t)$.
Then the desired spanning tree of $G$ is given by $T = (V, E_T)$.

\item
\begin{enumerate}
\item
For each tree edge $(u,v) \in E_T$, we compute $d(u)$, $d(v)$, $p(u)$ and $p(v)$ via an
Euler-Tour traversal of $T$. We also preprocess $G$ using the LCA algorithm 
given in Corollary~\ref{cor:LCA}.
\item
Number the edges of $G$, assigning each $e \in E$ an arbitrary integer serial number
$\cf{serial}(e) \in [E]$. 
Define a lexicographic order on the non-tree edges $f$ according to 
$\cf{number}(f) = \left(d(\lca(f)), \cf{serial}(f)\right)$
\end{enumerate}

\item 
Each non-tree edge $f$ induces a simple cycle in $(V, E_T \cup f)$.
We will adapt an algorithm of Vishkin~\cite{v-epso-85} to 
compute compute a function $\cf{master}(e)$
for each tree edge $e$. 
	
\begin{fact}\cite{msv-pedsstn-86}
Each non-tree edge $f$, together with the set of edges $e_i$ such that $\cf{master}(e_i) =
f$ form a simple path or cycle, called the \emph{ear} of $f$.
\end{fact}

\begin{fact}\cite{msv-pedsstn-86}
The lexicographical order on $\cf{number}(f)$ over the non-tree edges induces an order on
the ears, which yields an ear decomposition (which is not necessarily open).
\end{fact}

\begin{fact}\cite{msv-pedsstn-86}
Let $(u,v)$ be a non-tree edge. Let $x = \lca(u,v)$. Let $e_u$ and $e_v$ be the first
edges on the path from $x$ to $u$ and $v$ in $T$ respectively. 
Then, the ear induced by $(u,v)$ is closed if and only if $e_u$ and $e_v$ both choose
$(u,v)$ as their \cf{master}.
\end{fact}

As shown by Maon~\etal~\cite{msv-pedsstn-86}, the resulting ear-decomposition is
not necessarily open, and we must first refine the order defined on
the edges. 
Therefore, we refine the ordering on non-tree edges sharing a common LCA 
by updating the assignment of \cf{serial} with the following additional steps.

\item
Construct a bipartite graph $H_x = (V_x, E_x)$ for each vertex $x \in V - \{t\}$. 
Each vertex in $V_x$ corresponds to an edge in $G$. Specifically, $V_x$ is the set of 
edges $(u,v)$ such that $x = \lca(u,v)$. Note that this includes tree edges $(x, w)$
for which $x = p(w)$. Thus, the graphs $H_x$ partition the edges of $G$.

There is an edge in $E_x$ between a tree edge $e$ and a non-tree edge $(y,z)$ if
and only if $e$ is the first edge on the path from $x$ to $y$ or $z$ in $T$.
There are no other edges in $E_x$. 
We can test all such edges by creating a list of each endpoint in $V_x$ sorted by
preorder number. That is, all tree edges $(x, w)$ will appear once ordered by
$\cf{preorder}(w)$ and each non-tree edge $(y,z)$ will appear twice, once
ordered by $\cf{preorder}(y)$ and once ordered by $\cf{preorder}(z)$. Then, each
non-tree edge is incident to the first tree edge that comes before and after it
according to the above ordering. 

We compute the set $E_x$ for all the graphs $H_x$ ``in parallel''; 
that is, we do not
require separate passes over the edge list of $G$ to compute each $H_x$.
First, in $O(1)$ rounds of compressed scanning, we label each edge $e$ with
$\lca(e)$. Then, we sort the edge list by $\lca(e), \cf{E-order}(e)$. In an
additional $O(1)$ rounds of compressed scanning, we compute $V_x$ and $E_x$ for
each $H_x$.

\item
Construct a spanning forest for each $H_x$. Note that each edge in $E$ appears
in exactly one $H_x$, and that the degree of each non-tree node is at most $2$.
Hence, the total size of all such graphs $H_x$ is $\theta(E)$.  We construct all
the spanning forests in two Bor\r{u}vka-style rounds as follows.  In the first
round, each non-tree node in $V_x$ selects one of its incident edges.  This
results in a spanning forest.  Then in a cleanup phase, we contract each
connected component into a pseudo-node, and remove loops and duplicate edges
between pseudo-nodes.  In the second round, each non-tree node in $V_x$ selects
its second edge (if present). We perform a second cleanup phase. The remaining
selected edges form a forest consisting of a spanning tree over each connected
component of each $H_x$.  Clearly we can implement the selection and cleanup
phases in a constant number of compressed-scanning rounds.

\item
\begin{fact}\cite{msv-pedsstn-86}
For each connected component of each $H_x$, there exists at least one tree edge 
$e \in V_x$ such that $d(\lca(\cf{master}(e))) < d(x)$.
\end{fact}

In light of this fact, we do the following:
\begin{enumerate}
\item
For each connected component $C$ of each $H_x$, find such an edge $e$ guaranteed by the 
above fact and construct an Euler Tree $T_C$ over $C$ rooted at $e$.
\item
Compute the pre-order numbers \cf{preorder}$(e)$ of each non-tree edge $e$ 
with respect to $T_C$ by performing an
Euler tour traversal of $T_C.$
\item
Recall that the edges were ordered according to the number assigned above:
$\cf{number}(f) = \left(d(\lca(f)), \cf{serial}(f)\right)$.
We reorder the non-tree edges by replacing their old serial numbers with the new preorder
numbers as follows:
$\cf{newnumber}(f) = \left(d(\lca(f)), \cf{preorder}(f)\right)$.
\end{enumerate}

\item
Now we are ready to compute the assignment of $\cf{master}$ based 
on the new ordering of the non-tree edges.
We proceed as follows:
\begin{enumerate}
\item Compute $\cf{preorder}(v)$ and $\cf{postorder}(v)$ for each vertex $v \in V$ by an
Euler Tour traversal of $T$. 
\item
For each vertex $v$, let $E_v$ denote the set of non-tree edges incident to $v$. 
If $\lca(E_v) \neq v$, then let 
 $(u,v) \in E_v$ be an edge such that $\lca(u,v) = \lca(E_v)$. That is, the edge
 $(u,v)$ such that $d(\lca(u,v))$ is minimized. If there is more than one such edge
 for a given vertex, choose a single edge arbitrarily.
 
 Assign  a tuple 
 $\cf{serial}(u,v) = (d(\lca(u,v)),\cf{serial}(u,v))$ to
 each chosen edge $(u,v)$ combining the depth of the
 least common ancestor and its old serial number.
 Note that the edge $(u,v)$ can be found by
 scanning the adjacency list of $v$ in $E - E_T$ after the above LCA preprocessing.
 Thus, we can relabel the serial numbers of all chosen edges 
 in a single compressed-scanning pass. 

As shown by Vishkin~\cite{v-epso-85}, 
all the non-tree edges that were not chosen in the previous step can
be discarded for the remaining computation of $\cf{master}$.

\item
For each tree edge $e = (p(v), v)$, initialize $\cf{master}(e) = f$, where $f$ is
the non-tree edge incident to $v$ with minimum serial number among the edges chosen in the
previous step. Note that the initial assignment of \cf{master} can be computed for all
tree edges in a single scan after sorting 
the chosen edges and tree edges in adjacency list order. 

\item
Perform a bottom-up traversal computation on $T$.
For each tree edge $e = (p(v), v)$, we assign a new value for $\cf{master}(e) = g$, 
where $g$ is the edge with minimum \cf{master}$(g)$ 
among all the edges between $(p(v), v)$ and $(v, p(v))$ in the
Euler Tour traversal of $T$.

\end{enumerate}

\end{enumerate}

\begin{fact}\cite{msv-pedsstn-86}
The set of non-tree edges chosen as $\cf{master}$ partition the tree edges into the
subsets that chose them. Each \cf{master} edges induces an ear, and the ordering of the
corresponding non-tree edges results in an open ear decomposition of $G$.
\end{fact}

\ifFull
This fact, together with the above algorithm, give us the following theorem.
\fi

\begin{theorem}
\label{app:thm:ear-decomp}
Given a biconnected graph $G$, and a spanning tree of $G$, 
we can construct an open ear decomposition in $O(\sort{E})$ time in the DO-OEM
model.
\end{theorem}
\ifFull
\begin{proof}
There are a constant number of steps outlined above for computing the ear decomposition.
It was already shown by Maon~\etal~\cite{msv-pedsstn-86} that completing these steps
correctly yields a valid open ear decomposition. It remains to consider the time taken by
each step. 
Each step involves a constant number of compressed scanning
rounds and Euler tree traversals, which each take at most $O(\sort{E})$ time.
Therefore, the running time of the entire algorithm is $O(\sort{E})$. 
Since the algorithm essentially reduces to data-oblivious sorting and
compressed-scanning, no probabilistic polynomial adversary can have more than
negligible advantage in the input-indistinguishability game.
\end{proof}

\section{st-Numbering}
\label{sec:st-num}
At a high level, our algorithm for st-numbering is similar to the one given
in~\cite{msv-pedsstn-86}. That is, our algorithm is also structured in two
stages; in the first stage, we orient each ear in the ear decomposition based on
some simple rules, and in the second stage we compute the st-numbering by
traversing the ears in an order based on their orientations. 
Necessarily the details of each stage are significantly different in the
DO-OEM model. 

The input to our st-numbering algorithm is an open ear decomposition, structured
as a sequence of ears (edge disjoint paths) 
$P_i$, where $P_0 = (s,t)$. Each ear $P_i$ has a left and right endpoint denoted 
$L(P_i)$ and $R(P_i)$ respectively.
  Each vertex in $P_i$ which is not an
endpoint of $P_i$  is
called an \emph{internal} vertex of $P_i$.
Note that each vertex in an internal vertex in exactly one ear. We say that a
vertex is \emph{belongs to} an ear only if it is an internal vertex of that ear.
  \joe{I'm open to using phrasing other than ``belongs to'', but having a similar
  definition makes it easier to write about later}
  The endpoints of each ear
other than $P_0$ belong to two (not necessarily distinct) ears
which occur earlier in the
sequence. 

Since the algorithm numbers internal vertices based on the ear which contains
them, we can safely ignore any ears (other than $P_0$) which do not contain any
interior vertices. Thus, all single-edge ears other than $P_0$ are 
discarded, and the algorithm is only run on the remaining graph which has size
$O(V)$.
  
We assume that each edge is augmented with the index of the ear which
contains it and the index of the ear containing each endpoint. Furthermore each
edge has a pointer to adjacent edges in the path, indicating whether the
neighboring edge is towards $L(P)$ or $R(P)$. Finally, the internal vertices
which belong to
each ear are
numbered according to the left to right order from $L(P)$ to $R(P)$, and each edge
is augmented with the numbers of its endpoints. 
If needed, the augmentation can easily be computed with compressed-scanning
steps.

\subsection{Orientation}
The vertex $L(P)$ is also called the \emph{anchor} of ear $P$. There is a natural tree
structure given by the ears and their anchors. The \emph{ear tree} $ET$ is a
directed tree rooted at $P_0$. It is formally defined
as follows:
Each ear $P_i$ is a vertex in $ET$.
There is an edge $(P_j, P_i)$ for each pair of ears such that the anchor of
$L(P_i)$ of $P_i$ belongs to $P_j$.
We say that vertex $u$ belonging to $P_i$ is a descendant of a vertex $v$ belonging to
$P_j$ if $P_i$ is a descendant of $P_j$ in $ET$.

We construct $ET$ in a constant number of compressed scanning rounds. We number
each edge in $ET$ according to its order in a preorder traversal of $ET$ via an
Euler tour traversal. Then, we augment a set of non-tree edges $\{L(.), R(.)\}$. For
each ear $P_i$, we precompute $\lca(L(P_i),R(P_i))$ using the least
common ancestor of Corollary~\ref{cor:LCA}. 
Let $e_L$ and $e_R$ be the first tree
edge on the path from $ \lca(L(P_i),R(P_i))$ to $L(P_i)$ and $R(P_i)$ respectively. 
In a constant number of additional
compressed scanning rounds, we store with each non-tree edge $\{L(P_i), R(P_i)\}$ the
data, including pre-order number and left-to-right number, associated with $e_L$ and $e_R$.


For each ear $P_i$, the orientation of $P_i$ depends only on the orientation
previously assigned to some individual ear $P_m$, called the \emph{hinge} of
$P_i$. We say that an ear gets the \emph{same direction} as its hinge if either both
ears are oriented from $L(.)$ to $R(.)$ or both are oriented from $R(.)$ to $L(.)$.
Otherwise, we say the ears get \emph{opposite} directions.

The set of ears and their hinges also form a tree structure. The
\emph{hinge tree} $HT$ is formally defined as follows:
Each ear $P_i$ is a vertex in $HT$. There is an edge $(P_m, P_i)$ for each pair
of ears such that $P_m$ is the hinge of $P_i$.
The hinge of each ear $P_i$ can be determined by a few constant time tests based
on the information stored in $(L(P_i), R(P_i))$ during the above preprocessing
phase. Details of these tests is given is Section~\ref{sec:hinge}. 
Given the hinges, we can build the hinge tree in a constant number of
compressed scanning rounds. Finally, we can perform an Euler tour traversal of
the hinge tree in order to determine the orientation of each ear.

\subsection{Numbering}
An ear oriented from right to left (towards its anchor) is called an
\emph{incoming} ear, 
and an ear oriented from left to right (away from its anchor) is called an
\emph{outgoing} ear.
We define a total order on the ears based on their orientations.
The internal order of incoming ears is the same as the order of the corresponding edges
in $ET$, and the internal order of outgoing ears is opposite of the
corresponding edges in $ET$. Each incoming ear is before each outgoing ear.
Thus, we can compute the relative order of any two edges in constant time given
their orientations. Hence, in $O(1)$ rounds of compressed scanning, we can
compute the st-numbering, which yields the following theorem.

\begin{theorem}
\label{app:thm:st-numbering}
Given a biconnected graph $G$ and a spanning tree for $G$,
we can find an st-numbering of $G$ in $O(\sort{E})$
time in the DO-OEM model.
\end{theorem}
\begin{proof}
As outlined above, the algorithm consists of an orientation phase and a
numbering phase. We also need to perform some pre-processing of
least-common-ancestors in the ear tree. Each phase requires at most a constant
number of tree traversal computations. However, the
input to each phase has size $O(V)$, since the algorithm initially discards
any ears other than $(s,t)$ without internal nodes. Therefore, each phase has
running time $O(\sort{V})$ and the running time is dominated by the time
required to discard the uninteresting ears, which is $O(\sort{E})$.

Since the algorithm is composed of data-oblivious sorting and
compressed-scanning, no probabilistic polynomial adversary can have more than
negligible advantage in the input-indistinguishability game.
\end{proof}

\subsection{Hinge Finding}
\label{sec:hinge}
We now describe how to find the hinge of each ear. The following is based on the
case analysis of Maon~\etal~\cite{msv-pedsstn-86} and is included here for completeness.
There are no significant differences in the assignment of orientation, and thus
correctness follows.
The primary differences are the extra care we need to take to efficiently maintain the
data-oblivious property of our algorithm. This is handled by storing the
necessary data with each ear $P_i$ during the above preprocessing of $(L(P_i),
R(P_i))$. 

There are two cases two consider.
\begin{enumerate}
\item 
If $L(P_i)$ and $R(P_i)$  belong to the same ear $P_j$, then $P_j$ is the hinge
of $P_i$.  We orient $P_i$ in the same direction as $P_j$ if and only if
$L(P_i)$ is left of $R(P_i)$ in $P_j$.

Let $e_L$ and $e_R$ be the edges in $P_i$ incident to $L(P_i)$ and $R(P_i)$
respectively. In constant time, we can test if $L(P_i)$ and $R(P_i)$ belong to
the same ear by looking at the ear index of $L(P_i)$ and $R(P_i)$ stored in
$e_L$ and $e_R$. We can also test in constant time whether $L(P_i)$ is left of
$R(P_i)$ in $P_j$ by looking at the left-to-right numbering stored in $e_L$ and
$e_R$.

\item 
If $L(P_i)$ and $R(P_i)$ belong to different ears $P_j$ and $P_k$ respectively, 
then there are several cases to consider.  
Let $P_\alpha$ be the LCA of $P_j$ and $P_k$ in $ET$.
    \begin{enumerate}
        \item 
        If $P_\alpha$ is distinct from $P_j$ and $P_k$, then
        let $v_j$ and $v_k$ be the anchor ancestors of $P_j$ and $P_k$
        respectively. There are two cases to consider:
        \begin{enumerate}
            \item 
                If $v_j$ and $v_k$ are distinct (i.e. $v_j \neq v_k$),
                then $P_\alpha$ is the hinge of $P_i$. Similarly to case 1, we
                orient $P_i$ in the same direction as $P_\alpha$ if and only if
                $v_j$ is to the left of $v_k$ in $P_\alpha$.
            \item 
            If $v_j = v_k$, then let $v = v_j = v_k$, and let 
            $e_j = (v, P_{j'})$ and $e_k = (v, P_k')$ be the first edge on the path from
            $P_\alpha$ to $P_j$ and $P_k$ respectively. If $e_j$ comes before
            $e_k$ in the adjacency list from $v$, then the hinge of $P_i$ is
            $P_j'$ and $P_i$ gets a direction opposite to $P_j'$. Otherwise
            $P_k'$ is the hinge of $P_i$, and $P_i$ gets the same direction as
            $P_k'$. Given the information stored with $(L(P_i), R(P_i)$ during the LCA
            preprocessing, we can determine the hinge of $P_i$ and whether it
            gets the same or opposite direction of its hinge in constant time. 
        \end{enumerate}
        \item 
        If $P_\alpha = P_k$, then let $v$ be the anchor ancestor of $P_i$ in
        $P_\alpha$.
        \begin{enumerate}
        \item 
        If $v \neq R(P_i)$ then $P_\alpha$ is the hinge of $P_i$, and $P_i$ gets
        the same direction as $P_\alpha$ if $v$ is to the left of $R(P_i)$ in
        $P_\alpha$.
        \item 
        If $v = R(P_i)$, then the hinge of $P_i$ is set to the first ear
        on the path from $P_\alpha$ to $P_j$, and $P_i$ is given the
        opposite direction of that ear.
        \end{enumerate} 

    \end{enumerate}

\end{enumerate}
\fi

\section{Conclusion}
We provided several I/O-efficient algorithm for fundamental graph
problems in the DO-OEM model, which are more efficient than 
simulations of known graph algorithms using existing ORAM simulation
methods
(e.g., see~\cite{sss-tpor-12,gghjrw-oo-13,gmot-pos-12,ss-os-2013,sdsfryd-po-13}).
Moreover, our methods are based on new techniques and novel adaptations of 
existing paradigms to 
the DO-OEM model (such as our bottom-up and top-down tree computations).

\bibliographystyle {splncs03}
\bibliography {../go,cuckoo}

\clearpage
\begin{appendix}

\ifFull
\section{Euler Tour of a Tree}
\label{app:euler-tour}
Many classic tree algorithms are based on depth first search. However, it is
difficult to perform DFS efficiently in parallel. Therefore, 
\emph{Euler tours} were proposed as an algorithmic technique for parallel computations
on trees by Tarjan and Vishkin~\cite{tv-epba-85}, 
and later adapted for data-oblivious algorithms by Goodrich~\etal~\cite{got-gdc-12}.

Given an undirected rooted tree $T$, we imagine that each edge $\{p(v), v\}$ is
composed of two directed edges $(p(v), v)$ and $(v, p(v))$, called an
\emph{advance} edge and \emph{retreat} edge respectively. An \emph{Euler tour}
of $T$ visits these directed edges in the same order as they would be visited in
a depth first search of $T$. 
Let \cf{E-order}$(u,v)$ denote the order of the
edge $(u,v)$ in an \emph{Euler tour} of $T$. 
Note that 
$\cf{preorder}(v) = \cf{E-order}(p(v),v)$ and 
$\cf{postorder}(v) = \cf{E-order}(v,p(v))$. 

We assume that 
for each vertex, the outgoing edges are arranged in a circular linked list.
We also assume that each edge is stored as a pair of directed edges, with bi-directional
pointers between the members of each pair. 
Thus, each edge has a pointer to \cf{next}, \cf{prev}, and \cf{twin}.
If a vertex is a leaf, then the single incident edge has \cf{next} and \cf{prev} pointers
which just point back to itself.
If we do not initially have these pointers, it is easy to create them from an unordered
list of edges in a constant number of compressed-scanning rounds.

To perform the Euler tour, we start with an arbitrary edge \cf{first} from the root.
The following code will list all the edges in Euler Tour order.

\begin{verbatim}
output first
edge e = first.twin.next
while e != first
    output e
    e = e.twin.next
\end{verbatim}
For example, consider the following tree:

\begin{figure}[hb!]
\vspace*{-12pt}
\begin{center}
\includegraphics{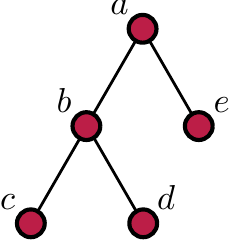}
\end{center}
\vspace*{-12pt}
\end{figure}

\noindent
If the first edge is $(a,b)$, then the edges are output:
\[
(a,b), (b,c), (c,b), (b,d), (d,b), (b,a), (a,e), (e,a).
\] 
Then $(e,a)\cf{.twin} = (a,e)$, 
 and $(a,e)\cf{.next} = (a,b)$ which is the first edge, so the loop terminates.

Some tree statistics are straightforward to compute
using Euler Tours. For example, Goodrich~\etal~\cite{got-gdc-12} show how to
compute the size of the subtree for each node $v \in T$ using an Euler Tour
and compressed-scanning pass over the edges of $T$. 
The calculation is straightforward once we observe that
$\cf{size}(v) =  (\cf{E-order}(v,p(v)) - \cf{E-order}(p(v), v))/2 + 1$,
since for each proper descendant of $V$, we will traverse one advance edge and
one retreat edge. Thus, the number of edges traversed between $(p(v), v)$ and
$(v, p(v))$ is twice the number of proper descendants of $v$, and we add one to
also include $v$ in $\cf{size}(v)$. Also note that the size of the root of $T$ is $n
= max(\cf{E-order}) / 2 + 1$.
\fi

\ifFull
\else
\ifSim
\input{sim}
\fi
\fi

\end{appendix}

\end{document}

%% file: intro.tex
\section{Introduction}
\ifFull
\emph{Outsourced external memory} is a computational framework
where a client performs computations on data
stored at a semi-trusted server.
This approach facilitates reliability 
over traditional frameworks, but it also introduces a challenge
with respect to privacy, since the server has full access to the data and he can
monitor how it is accessed by the client.
That is,
a client outsources her data to an external server administered by a
third party so that she can reliably access her data from anywhere using any
computational device.
Moreover, the client gains these
features of reliability and availability often at very low cost (sometimes
they are even free).
Unfortunately, this approach introduces a
loss of privacy that can occur from outsourcing data to a third party.
Indeed, some cloud storage companies have based their business models on
their ability to mine client data for 
useful information.
Even if data is encrypted, information can be leaked from the way it is
accessed. 
For example, Chen {\it et al.}~\cite{cwwz-sclwa-10} are
able to infer sensitive information from the access patterns of popular
health and financial web sites even if the data streams are encrypted.
Thus, it is useful to design methods that allow for privacy-preserving
access to data in the cloud.
We are therefore interested in privacy-preserving algorithms
for outsourced external memory
that are based on the use of data-oblivious algorithms, that is,
algorithms where each possible sequence of data accesses 
is independent of the data values.
Such algorithms are useful for computation on outsourced data, 
since combining them with a
semantically-secure encryption scheme 
will not reveal data values nor data access patterns.
\fi

In this paper, 
we work within the 
\emph{data-oblivious outsourced external memory} (DO-OEM) model,
which is our name for the model used in recent papers on algorithms
and systems for data-oblivious outsourced storage solutions
\ifFull
(e.g., see~\cite{MIT-CSAIL-TR-2011-018,sss-tpor-12,gghjrw-oo-13,gmot-pos-12,gkkkmrv-stpc-12,scs-or-11,ss-os-2013,sdsfryd-po-13,wss-poos-11}).
\else
(e.g., see~\cite{sss-tpor-12,gghjrw-oo-13,gmot-pos-12,ss-os-2013,sdsfryd-po-13}).
\fi
We assume that a large data set of size $N$ is stored on a server, 
who we will call ``Bob,''
and that a client, ``Alice,''
has access to this data through an I/O interface that allows her to
make read and write requests of Bob using messages of size $B$ as atomic actions.
We also assume
Alice has a small amount of secure, private working memory,
of size $M = \Omega(\log N)$.

The server, Bob, is ``honest-but-curious,''
which means that he will correctly perform 
every task requested, but he will also try to learn as much as possible about
Alice's data.
This, of course,
introduces privacy 
constraints for the DO-OEM model 
not found in the traditional I/O model (such as in~\cite{Chiang:1995}).
\ifFull
In particular, we can rely on Bob to faithfully execute read and
write requests, but he would like to learn as much as possible about Alice's
data.
\fi
Thus, Alice must encrypt her data and then decrypt it and re-encrypt it with each
read and write request, using a semantically-secure encryption scheme.
Alice can safely perform any computation in her private memory, 
but her sequence of data accesses on the server must also not leak
information about her data. 
That is, it must be \emph{data oblivious}.
The access sequence may depend on the function being
computed, but it should be independent 
of the input data values. 

Formally, we suppose Alice wants to perform an algorithm, $A$,
which computes some function, $f$, on her data stored with Bob.
In the context of graph algorithms, the input to $f$ is a graph, usually
formatted as an array of edges, with $V$ and $E$ being the number
of the graph's vertices and edges, respectively. 
The output of $f$ may either be a
property of the graph, such as whether or not the graph is biconnected,
or another graph, such as a spanning tree, which will also be stored
with Bob.
Alice performs the algorithm $A$ by issuing read and write requests to Bob. 

We say that $A$ is \emph{data-oblivious} and can compute $f$ in the DO-OEM
model if every probabilistic polynomial time adversary 
has only a negligable advantage over random guessing in a 
\emph{input-indistinguishability} game.
\ifFull
\begin{enumerate}
\item 
The challenger sends a description of $f$ and $A$ to the adversary.
\item 
The adversary performs some computation, possibly
including simulating $A$ on a polynomial number of inputs.
The adversary chooses some input parameters and chooses an input satisfying the
parameters. 
The adversary sends the input array and input parameters to the challenger. 
\item The challenger generates an input satisfying the input parameters uniformly at random.
The challenger flips a fair coin to decide whether to use the input provided by
the adversary or to use the randomly generated input array. The challenger
encrypts the selected input, and writes the encrypted array to a shared location. 
\item
The challenger runs algorithm $A$ on the shared encrypted array.
\item
The adversary observes the sequence of reads and writes to the encrypted array,
but does not see the challenger's private memory or the unencrypted values of
data in the array. 
After polynomial computation, the adversary guesses
which input was used by the challenger. The adversary wins the game if it
guesses correctly, otherwise the challenger wins.
\end{enumerate}
\else
In this game, Alice flips a fair coin and based on its outcome either
uses $A$ to compute $f$ on her input or on a random syntactically-correct input.
Bob observes her memory accesses and then must decide if she was
computing $f$ for her input or not.
\fi
Given a function $f$ and public input parameters
(e.g. an upper bound on the size of the graph),
$\gamma$, 
the probability that an algorithm to compute $f$
in the DO-OEM model executes a particular access sequence $S$
must be equally likely for any two inputs 
$X,Y$ satisfying parameters $\gamma$. That is,
$
P(S | f,\gamma, X) = P(S | f, \gamma, Y)
$,
or, from the Bob's perspective, 
$ P(X | f,\gamma,S) = P(Y | f,\gamma, S)$. 
We can achieve data-obliviousness if,
knowing the size of the input, the function being computed, and the
access sequence, all inputs are equally likely.

\ifFull
Moreover, for the problems we study in this paper, 
we would like to avoid using
constant-time random oracles, since the existence of 
such functions is considered a strong assumption in 
the cryptographic literature (e.g., see~\cite{dmn-psorwro-11}).

We use notation similar to the 
standard external memory model~\cite{v-emads-01}, 
but standard external memory techniques will not lead to
data oblivious algorithms. Thus, although we measure the running time of our
algorithm in I/Os, we require novel techniques in order to achieve
data oblivious algorithms.
\fi

\paragraph{Previous Related Results.}
\ifFull
Oblivious algorithms are discussed in a classic book
by Knuth~\cite{k-ss-73}, and
Pippenger and Fischer~\cite{pf-racm-79} 
show how to simulate a one-tape Turing machine of length $n$ with
an oblivious two-tape Turing machine computation of length $O(n\log n)$.
\fi
Goldreich and Ostrovsky~\cite{go-spsor-96} introduce
the oblivious RAM model and show that an arbitrary RAM algorithm
can be simulated (in internal memory) with an overhead of $O(\log^3 N)$ through the use
of constant-time random oracles, and this has subsequently been 
improved 
(e.g., see~\cite{gm-ppaod-11,gmot-ors-11,Goodrich:2012,gmot-pos-12}),
albeit while still using constant-time random oracles.
\ifFull
Ajtai~\cite{a-orwca-10} shows how to perform oblivious RAM simulation 
with a polylogarithmic factor overhead 
without constant-time random oracles and
\fi
Damg\aa{}rd {\it et al.}~\cite{dmn-psorwro-11} show how
to perform such a simulation with an $O(\log^3 N)$ overhead without
using random oracles.

Chiang {\it et al.}~\cite{Chiang:1995} study (non-oblivious) external-memory
graph algorithms and
Blanton {\it et al.}~\cite{bsa-doga-13} give data-oblivious algorithms
for breadth-first search, single-source-single-target shortest paths and minimum
spanning tree with running time $O(v^2)$, 
and maximum flow with running time $O(v^3 E \log V)$. However, their approach is based on
computations on the adjacency matrix of the input graph, and thus only optimal
on very dense graphs, whereas our approach is based on reductions to sorting the
edge list, and is efficient on graphs of all densities.

\ifFull
Finding
minimum spanning trees is a classic, well-studied
algorithmic problem with many applications.
Likewise,
computing an
st-numbering is vital in a number of graph drawing and planarity testing
algorithms, and st-numbering in the data-oblivious model was listed as an open
problem~\cite{got-gdc-12}.
Thus, designing efficient
DO-OEM algorithms for these problems can result in improved
privacy-preserving algorithms for a number of other problems.
\fi


\paragraph{Our Results.}
Let \sort{N} denote the number of I/Os required to sort an input of
size $N$ in the DO-OEM model.
For instance,
Goodrich and Mitzenmacher~\cite{gm-ppaod-11}
show
$\sort{N} = O((N/B) \log_{M/B}^2 (N/B))$ 
I/Os, assuming $M > 3B^4$.
We develop efficient algorithms in the DO-OEM
model for a number of fundamental graph problems:
\begin{itemize}
\item
We show how to
construct a minimum spanning tree of a graph  $G$ in 
time depending on the input parameters $V$, $E$, density, and class of $G$,
in the DE-OEM model:
\ifFull
If $G$ belongs to
a minor-closed family of graphs, such as a planar graph or any graph with bounded genus, 
then the run-time is $O(\sort{E})$.
\fi
\begin{center}
\begin{tabular}{|c|c|c|c|}
\hline
Density & Class & Running Time & Constants \\  \hline
$E = O(V \log^\gamma V)$ & Any & $O(\sort{E} \log  V / \log \log V)$ & $\gamma \geq 0$ \\ \hline
$E = \Theta(V2^{\log^\delta V})$ & Any & $O(\sort{E} \log^{1-\delta}V)$ & $0 < \delta < 1 $  \\\hline
$E = \Omega(V^{1+\epsilon})$ & Any &  $O(\sort{E})$ & $0 < \epsilon \leq 1 $ \\ \hline
Any & Minor Closed & $O(\sort{E})$ & --- \\\hline
\end{tabular}
\end{center}
\item 
Given a tree $T$, we can perform any associative 
traversal computation over $T$ in $O(\sort{V})$ time in the DO-OEM model.
\item
Given a tree $T=(V,E)$ and a set, $S$, of 
pairs of vertices,
we can compute 
the LCA\ifFull,
the least common ancestor,\fi{} for each pair in $S$ in $O(\sort{|S| + V})$ 
time in the DO-OEM model.
\item
Given a graph $G$ and a spanning tree of $G$, we can compute the biconnected
components of $G$ in $O(\sort{E})$ time in the DO-OEM model.
\item
Given a biconnected graph $G$, and a spanning tree of $G$, 
we can construct an open ear decomposition in $O(\sort{E})$ time in the DO-OEM
model.
\item
Given a biconnected graph $G$ and its open ear decomposition,
we can find an st-numbering of $G$ in $O(\sort{E})$
time in the DO-OEM model.
\ifSim
\item
Given CRCW PRAM algorithm $A$ that runs in $t$ steps using
a memory of size $N$ and $P\le N$ processors,
we can simulate $A$ sequentially in the DO-OEM model in $O(t\cdot\sort{N})$ time. 
Due to space constraints, this result is
given in Appendix~\ref{sec:sim}
\fi
\end{itemize}
None of our algorithms use constant-time random oracles. Instead,
they are based on a number of new algorithmic 
techniques and non-trivial adaptations of existing techniques.
\ifFull\else
This paper outlines the main results; details can be found in the full
version~\cite{GoS14}.
\fi

\section{Data-Oblivious Algorithm Design}
\label{sec:prelim}
\ifFull
\subsection{Preliminaries}
As is common in external memory graph algorithms we denote the input graph 
$G=(V,E)$, and slightly abuse notation by letting $V = |V|$ and $E = |E|$ when
the context is clear. All logs are base 2 unless otherwise indicated. 
We use $[V]$ to denote the set of integers in the range
$[1, |V|]$, and use $T$ to denote a rooted spanning tree of $G$.
For each $v \in V$, $d(v)$ denotes the \emph{depth} of $v$, i.e. the distance from the
root in $T$. The parent of each node with respect to $T$ is denoted by $p(v)$.
Let $T_v$ denote the subtree rooted at $v$, 
\cf{size}$(v)$ denote the number of vertices in $T_v$, and 
\cf{preorder}$(v)$ and \cf{postorder}$(v)$ denote the 
order of $v$ in a preorder and postorder
traversal of $T$ respectively.  
For each pair of nodes $u,v \in V$, $\lca(u,v)$ denotes the least common ancestor of $u$
and $v$ in $T$. 

Given a graph $G = (V,E)$ and an edge $(s,t) \in E$, an \emph{st-Numbering} of 
$G$ is a function $\ell: V \to [V]$ which assigns integer labels to $V$ such 
that $s$ is labeled $\ell(s) = 1$,
$t$ is labeled  $\ell(t) =V$, and every other vertex $v$ is adjacent to two vertices $u$
and $w$ such that $\ell(u) \leq \ell(v) \leq \ell(w)$.

An \emph{ear decomposition} of $G$ is a partition of $E$ into a set of simple paths
$P_0, P_1, \ldots P_k$ called \emph{ears} such that $P_0 = (s,t)$ and for each ear $P_j$, the
internal vertices are disjoint from all $P_i$ where $i < j$, but each endpoint 
$L(P_j)$ and $R(P_j)$ of $P_j$ is either contained in $P_0$ or the internal
vertices of some ear $P_i$, $0 < i < j$. 
An ear $P_i$ is \emph{open} if $L(P) \neq R(P)$ and an ear decomposition is
\emph{open}
if all its ears are open.

We assume that the input is formatted as a list of $V$ vertices and
$E$ directed edges sorted in adjacency list order. 
Note that we can organize an arbitrary
list of edges into this order with $O(1)$ rounds of compressed-scanning. 
Although the edge list does not have separate storage for the vertices of the
graph, for brevity, we sometimes say that we will store a data value at a vertex.
However, to associate any data with a vertex, we must store that data in each
edge incident to the vertex.  It requires $O(1)$ rounds of compressed-scanning 
to distribute this data to all the incident edges. 

\ifFull
\subsection{ORAM Simulation} 
There are existing methods
for simulating any random-access machine (RAM) algorithm 
    (in internal memory) in a data-oblivious way  
without the use of random oracles. 
The best such oblivous RAM (ORAM) simulation method~\cite{dmn-psorwro-11}
which avoids constant-time random oracles,
costs $O(\log^3 N)$ per memory access, however.
Thus, it is ideal if we can avoid
using such general simulation results as much as possible.
Therefore, in this paper, we are interested in methods for solving fundamental
graph problems directly in the DO-OEM model.
\fi

\subsection{Compressed Scanning}
\fi
\paragraph{Compressed-Scanning.}
\label{sec:sort-scan}
Goodrich~\etal~\cite{got-gdc-12}
introduced \emph{compressed-scanning} as a data-oblivious algorithm design technique for internal
memory. 
\ifFull
Another related model is the massive, unordered, distributed (MUD) 
model for map-reduce~\cite{fmsss-dssc-10}. 
\fi
We extend compressed-scanning to the DO-OEM model, and 
we use an algorithm design technique, where an algorithm
is formulated so that it processes the input in a series of $t$ rounds
as follows:
\begin{enumerate}
\item 
Scan each item of input exactly once; a random
permutation 
hides the access pattern.
\begin{itemize}
\item Read a block of $B$  
items from input, possibly including some dummy items.
\item Perform some computation in private memory.
\item Write a block of $B$  
items to output, possibly including some dummy items.
\end{itemize}
\item Sort the output data-obliviously.
\item Truncate the output, ignoring a portion, $L(i, \gamma)$, of the items, 
    which may depend on the index of the round $i$ and input parameters $\gamma$ (e.g.
    input size), but not on any data values. For example, $L$ could be $0$ (no
    items are discarded) or $N/2^i$ (the last half of the output
    is discarded in each round).
\item Use the output as input for the next round.
\end{enumerate}
\ifFull
This pattern is similar to the streaming model augmented with a 
sorting primitive~\cite{adrr-smasp-04}. 
\fi

\begin{theorem}
\label{thm:one}
 Let $A$ be any compressed-scanning algorithm
for which $t$ and $B$ depend only on $N$. Let $N_i$ denote the size of the input
passed to round $i$. Then, $A$
can be simulated in the  DO-OEM model in
$O(\sum_{i=1}^t \sort{N_i})$ time\footnote{ 
    Thoughout this paper we measure time 
    in terms of I/Os with the server.}
without the use of
constant-time random oracles.
\end{theorem}
\begin{proof}
    By definition, we can run $A$ in the DO-OEM model if and only if $A$
    satisfies the input-indistinguishability game. The algorithm runs in $t$
    rounds, and each round has three phases: scan, sort, and truncate. The adversary
    can win the game if in any round, in any phase, the sequence of memory
    accesses to the shared encrypted array is different with non-negligible 
    probability between the adversary's input 
    and input generated uniformly at random. 
    However, by construction, the distribution of memory access by $A$ is the same at
    every phase for all inputs with the same input parameters. 
    In the scan phase, each item of input is accessed once and in random order,
    regardless of the actual input values. Thus the scan phase conveys no
    advantage to the adversary. 
    In the sorting phase, the sequence of
    memory access is likewise independent of the data values by definition,
    since we use a data-oblivious sorting algorithm. Hence, the sorting phase
    also conveys no advantage to the adversary. 
    Finally, in the truncate phase, the portion of memory the algorithm chooses
    to ignore depends on the input parameters, but will be identical, regardless
    of whether the input is the one chosen by the adversary or the one
    generated uniformly at random. So the truncate phase also conveys no
    advantage to the adversary. In every phase of every round, the adversary
    gains no information as to which input was chosen by the coin toss. 
    Therefore,  every probabilistic polynomial time adversary has 
    a negligable advantage over random guessing.
    The running time is a straightforward sum of the cost to data-obliviously
    sort the input in each round. 
\end{proof}

\ifFull
As we explore in this paper,
using this design
approach results in much improved running times over 
a general RAM-simulation approach.
Thus, we 
would like to use compressed-scanning algorithms as much as possible.
This approach introduces some interesting challenges from an 
algorithmic perspective,
and designing efficient 
compressed-scanning algorithms for even well-known problems
often involves new insights or techniques (e.g., see~\cite{got-gdc-12}).
\fi